\documentclass[journal]{IEEEtran}
\usepackage{makecell}
\usepackage{array}
\usepackage{graphicx,amssymb,amsmath}
\usepackage{multicol}
\usepackage[noadjust]{cite}
\usepackage{setspace}
\usepackage{subfigure}
\usepackage{graphicx}
\usepackage{float}
\usepackage {url}
\usepackage{stfloats}
\usepackage{amsthm,pifont}
\usepackage{flushend}
\usepackage{cases,subeqnarray}
\usepackage{bm,multirow,bigstrut}
\usepackage{amsmath, amsthm, amssymb}
\usepackage{textcomp}
\usepackage{latexsym,bm}
\usepackage{booktabs}
\usepackage{xcolor}
\usepackage{epstopdf}
\usepackage{mathtools}
\usepackage{dsfont}
\usepackage{extarrows}
\usepackage{epsfig}
\usepackage{epsfig}
\usepackage{epstopdf}
\usepackage[noend]{algpseudocode}
\usepackage{algorithmicx,algorithm}
\renewcommand{\algorithmicrequire}{\textbf{Input:}}
\renewcommand{\algorithmicensure}{\textbf{Output:}}

\usepackage{cite}
\usepackage{bm}
\usepackage{cleveref}
\usepackage{multicol}       
\usepackage{multirow}       
\usepackage{array}          
\usepackage{colortbl}
\usepackage{makecell}
\definecolor{crimson}{RGB}{192,0,0}         
\definecolor{navy}{RGB}{47,85,151}         
\usepackage{bbding}
\usepackage{graphicx}
\usepackage{booktabs}
\usepackage{algorithm}
\usepackage{algpseudocode}

\theoremstyle{plain}

\theoremstyle{plain}

\usepackage{amsmath}

\IEEEoverridecommandlockouts

\begin{document}
\title{Multi-agent Reinforcement Learning-based Joint Precoding and Phase Shift Optimization for RIS-aided Cell-Free Massive MIMO Systems}

\author{{Yiyang Zhu, Enyu Shi, Ziheng Liu, Jiayi Zhang,~\IEEEmembership{Senior Member,~IEEE}, Bo Ai,~\IEEEmembership{Fellow,~IEEE}}
\thanks{This work was supported in part by the Fundamental Research Funds for the Central Universities under Grants 2023YJS015 and 2022JBQY004, in part by National Natural Science Foundation of China under Grants 62221001, in part by Natural Science Foundation of Jiangsu Province, Major Project under Grant BK20212002, and in part by ZTE Industry-University-Institute Cooperation Funds under Grant No. IA20240319002. (\emph{Corresponding author: Jiayi Zhang}).
}
\thanks{Y. Zhu, E. Shi, Z. Liu, J. Zhang, and B. Ai are with the School of Electronic and Information Engineering, Beijing Jiaotong University, Beijing 100044, P. R. China. (e-mail: jiayizhang@bjtu.edu.cn).}
}
\maketitle
 
\begin{abstract}
Cell-free (CF) massive multiple-input multiple-output (mMIMO) is a promising technique for achieving high spectral efficiency (SE) using multiple distributed access points (APs). However, harsh propagation environments often lead to significant communication performance degradation due to high penetration loss. To overcome this issue, we introduce the reconfigurable intelligent surface (RIS) into the CF mMIMO system as a low-cost and power-efficient solution. In this paper, we focus on optimizing the joint precoding design of the RIS-aided CF mMIMO system to maximize the sum SE. This involves optimizing the precoding matrix at the APs and the reflection coefficients at the RIS. To tackle this problem, we propose a fully distributed multi-agent reinforcement learning (MARL) algorithm that incorporates fuzzy logic (FL). Unlike conventional approaches that rely on alternating optimization techniques, our FL-based MARL algorithm only requires local channel state information, which reduces the need for high backhaul capacity. Simulation results demonstrate that our proposed FL-MARL algorithm effectively reduces computational complexity while achieving similar performance as conventional MARL methods. 
\end{abstract}


\IEEEpeerreviewmaketitle

\section{Introduction}


The sixth-generation (6G) network will be a vital component in all parts of future society, industry, and life, given its primary mission to fulfill the communication needs of humans and intelligent machines \cite{zhang2020JSAC}. The integration of distributed networks and massive MIMO confers notable advantages upon an ultra-dense network known as cell-free (CF) massive multiple-input multiple-output (mMIMO). In the context of CF mMIMO networks, a substantial array of distributed access points (APs) collectively cater to a limited user base using concurrent time-frequency resources, while all base stations (BSs) are linked to a central processing unit (CPU) through backhaul wireless connections \cite{9940169,chen2021structured}.


To enhance the network capacity, the deployment of a large number of distributed APs in the cell-free network is necessary. However, this approach entails significant costs and power consumption. Fortunately, the authors of \cite{shi2023ris} present a promising solution for enhancing network capacity in a cost-effective and energy-efficient manner, which involves using reconfigurable intelligent surfaces (RIS) to assist CF mMIMO systems. RIS is increasingly recognized as a forward-looking smart radio technology for advancing future 6G communications \cite{liu2021Star}. Consequently, the utilization of RIS-aided CF mMIMO systems can lead to improvements in channel capacity, reduced transmission power, enhanced transmission reliability, and expanded wireless coverage \cite{10225319,10326460,zhang2021ris,10264149,10290997}.

The utilization of joint precoding in RIS-aided CF mMIMO systems, as opposed to conventional precoding at the APs alone, involves the coordinated design of the beamforming matrix at the AP and the phase shifts of the RIS elements. This approach has been explored in recent research, such as the joint active-and-passive precoding framework proposed in \cite{9459505} and the partially connected CF mMIMO framework presented in \cite{10058895}. Additionally, efforts towards addressing practical implementation challenges of these techniques, such as the creation of less complex iterative algorithms, have been undertaken, as explored in \cite{10129196}. Despite these advancements, several challenges persist, particularly in the application of computationally intensive algorithms and joint learning in practical scenarios, necessitating further resolution for real-world deployment.

As a crucial technology for future 6G-and-beyond wireless communication systems, machine learning/artificial intelligence holds the promise of resolving non-convex optimization problems that are mathematically insoluble \cite{9775110}. Specifically, the authors in \cite{10278860} proposed a meta reinforcement learning (meta-RL)-based computation offloading policy to optimize RIS phase shift. In \cite{10296835}, the authors introduced a distributed machine learning-based approach to optimize the transmit beamforming at the AP. The aforementioned methods solve the computation complexity problems, however, the acquisition of instantaneous global CSI still entails substantial front-haul overhead \cite{liu2023cell}.

To address the challenges in the RIS-aided CF mMIMO system as mentioned above, inspired by multi-agent reinforcement learning (MARL), in this paper, we introduce an innovative MARL-based downlink design for joint precoding and phase shift to mitigate these challenges. The principal contributions of this paper are delineated as follows:

\begin{itemize} \item We investigate an RIS-aided CF mMIMO network and formulate the optimization problem for joint precoding and phase shift to maximize the sum-SE. In contrast to centralized training centralized execution approach, such as alternating optimization (AO), we employ the method of centralized training and distributed execution of MARL to solve this problem.

Based on the proposed MARL algorithm, we design a two-layer network to address the joint optimization of AP precoding and RIS phase shift separately. Each AP in our approach only requires local CSI for precoding design, reducing the burden and overhead on the backhaul link. Additionally, we introduce fuzzy logic (FL) to enhance the convergence speed of MARL and further reduce computational complexity. The results demonstrate that our proposed algorithm outperforms AO-based precoding regarding SE performance in a limited training time.

\end{itemize}

\textit{Notation}: The mathematical notation $(\cdot)^{H}$ is employed to denote the conjugate transpose operation. Boldface uppercase letters such as $\mathbf{X}$ are utilized to represent matrices, while boldface lowercase letters such as $\mathbf{x}$ are employed to denote column vectors. Furthermore, the complex Gaussian random variable $x$ with variance $\sigma^2$ is represented by $x \sim \mathcal{C}\mathcal{N}\left({0,{\sigma^2}} \right)$.

\begin{figure*}[ht]
\centering
\includegraphics[scale=0.45]{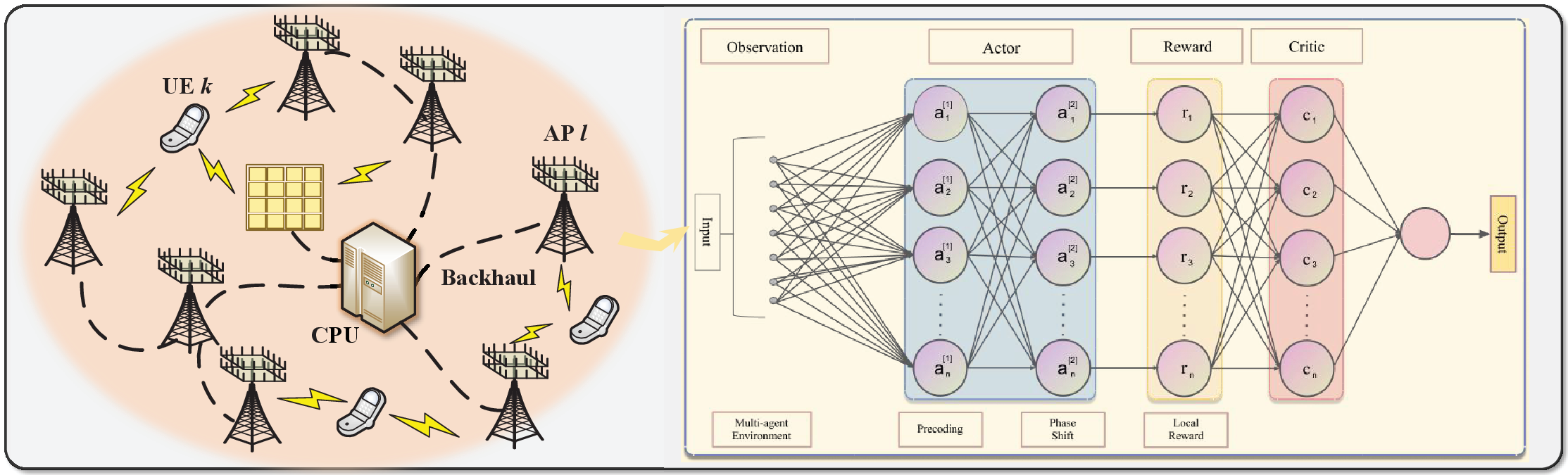}
\caption{The RIS-aided CF mMIMO system and the proposed MARL precoding network.\label{Fig1}} 
\end{figure*}

\section{System Model}\label{se:model}
In this section, we focus on an RIS-aided CF mMIMO system, as represented in Fig. 1, which utilizes several distributed APs and RISs to serve all UEs, simultaneously. For centralized control and training, a CPU is employed. All APs are connected to the CPU by optical cables or wireless fronthaul/backhaul links \cite{zhang2021improving,shi2023survey}. This design enables the distributed APs to obtain important user-specific CSI and cooperatively service all UEs. The control of all RISs is overseen by the CPU, and facilitated by wired connections. 

Specifically, we assume the network consists of \textit{L} APs, \textit{K} UEs, and \textit{R} RISs. We assume that each AP and UE are equipped with \textit{M} antennas, \textit{U} antennas, respectively. Also, each RIS consists of \textit{N} elements. We use the sets $\mathcal{N} = \{1,2,\dots,N\}$, $\mathcal{L} = \{1,2,\dots,L\}$, $\mathcal{K} = \{1,2,\dots,K\}$, and $\mathcal{R} = \{1,2,\dots,R\}$ to represent the index sets for RIS elements, APs, UEs, and RISs, respectively.
 
\subsection{Channel Model}
The utilization of RISs enables directional reflection, thereby structuring the channel between each AP and UE into distinct constituents. Specifically, the channel comprises an AP-UE link and \textit{R} AP-RIS-UE links, with each AP-RIS-UE link further divisible into an AP-RIS link and a RIS-UE link. The architectural framework, which is facilitated by RISs, delineates the communication channels within wireless systems, offering enhanced control and optimization of signal propagation.

A phase shift matrix to the incident signal, followed by the transmission of the phase-shifted signal to the user, is applied to represent on the RISs. Consequently, the resulting equivalent channel, denoted as $\mathbf{h}_{l,k}^{H}$, originating from the \textit{l}-th AP to the \textit{k}-th UE, is represented as
\begin{align}
    \mathbf{\hat{h}}_{l,k}^{H} = \mathbf{H}_{l,k}^{H} + \sum_{r=1}^{R} \mathbf{F}_{r,k}^{H}\mathbf{\Theta}_{r}^{H}\mathbf{G}_{l,r}^{},
\end{align}
where $\mathbf{G}_{l,r}^{} \in \mathbb{C}^{N\times M}$, $\mathbf{F}_{r,k}^{H} \in \mathbb{C}^{U\times N}$ denote the frequency-domain channel from the AP \textit{l} to RIS \textit{r}, and from RIS \textit{r} to UE \textit{k},respectively; $\mathbf{\Theta}_{r}^{H} \in \mathbb{C}^{N\times N}$ denotes the phase shift matrix as the RIS \textit{r}, which is written as
\begin{align}
    \mathbf{\Theta}_{r}^{H} \triangleq \text{diag}(\theta_{r,1},\dots,\theta_{r,N}),\forall r \in \mathcal{R},
\end{align}
where $\theta_{r,n}\in \mathcal{F}$. Note that $\mathcal{F}$ is the feasible set of the reflection coefficient at RIS. For simplicity but without loss of generality, here we assume $\mathcal{F}$ is the ideal case, i.e.,
\begin{align}
    \mathcal{F} \triangleq \{\theta_{r,n}\vert \theta_{r,n} \vert \leq 1\}, \forall r \in \mathcal{R}, \forall n \in \mathcal{N}.
\end{align}
Besides, $\mathbf{H}_{l,k}^{H} \in \mathbb{C}^{U\times M}$ denote the frequency-domain channel from the AP \textit{l} to the UE \textit{k}, which can be written as
\begin{align}
    \mathbf{H}_{l,k}^{H} = \beta_{l,k}^{H}\vert \mathbf{h}_{l,k}^{H} \vert ^{2},
\end{align}
where $\beta_{l,k}^{H}$ denotes the large-scale factor, $\mathbf{h}_{l,k}^{H}$ is the Rayleigh fading vector composed of the small-scale fading coefficients between AP \textit{l} and UE \textit{k}.

\vspace{-0.3cm}
\subsection{Transmitters and Receivers}
Our proposed RIS-aided CF mMIMO system establishes synchronization among all APs, a prerequisite for facilitating coherent joint transmission to cater to all users. Let $\mathbf{s} \triangleq [s_{1},s_{2},\dots,s_{K}]^{T} \in \mathbb{C}^{K}$ denote the vector of symbols, where each $s_{k}$ corresponds to the symbol transmitted to the \textit{k}-th user. It is ensured that the transmitted symbols adhere to power normalization, implying that $\mathbb{E}\{ss^{H}\} = \mathbf{I}$, with $\mathbf{I}$ denoting the identity matrix.

In the downlink, the frequency-domain symbol $s_{k}$ undergoes precoding using the precoding matrix $\mathbf{w}_{l,k} \in \mathbb{C}^{M}$ at the \textit{l}-th AP. The initial precoding operation yields the precoded symbol $\mathbf{x}_{l}$ and can be mathematically represented as
\begin{align}
    {{\bf{x}}_l} = \sum\limits_{k = 1}^K {{{\bf{w}}_{l,k}}{{\bf{s}}_k}}.
\end{align}

Let's represent the baseband frequency-domain signal received by UE \textit{k} as $\mathbf{y}_{k}^{}\in \mathbb{C}^{U}$
\begin{equation}
\begin{aligned}
        \mathbf{y}_{k}^{} &= \sum_{l=1}^{L}\mathbf{\hat{h}}_{l,k}^{H}\mathbf{x}_{l}^{} + \mathbf{z}_{k}^{H}.
\end{aligned}
\end{equation}

\vspace{-0.3cm}
\subsection{Problem Formulation}
Based on the system model above, the aim in this subsection is to enhance the overall SE gain realized over the network's operational duration.
At first, the signal-to-interference-and-noise ratio (SINR) for the transmitted symbol $s_{k}$ at UE \textit{k} is calculated as
\begin{equation}
\begin{aligned}
        \gamma_{k}^{} = \frac{\vert \sum_{l=1}^{L}\mathbf{\hat{h}}_{l,k}^{H}\mathbf{w}_{l}^{}\vert ^{2} }{\sum_{j \ne k}\vert \sum_{l=1}^{L} \mathbf{\hat{h}}_{l,k}^{H}\mathbf{w}_{j}^{}\vert ^{2}+\sigma^{2}}.
\end{aligned}
\end{equation}

Thereby, the SE of UE \textit{k} $R_{k}^{}$ is given by
\begin{equation}
    \begin{aligned}
        R_{k}^{} = \text{log}_{2}(1+\gamma_{k}^{}).
    \end{aligned}
\end{equation}

Finally, the optimization problem of maximizing SE gain can be originally formulated as
\begin{equation}
\begin{aligned}
\mathcal{P}^{0} \mathop {\max}\limits_{{{\bf{w}}_k},\Theta } \text{sum-SE} &= \sum\limits_{k = 1}^K {{{\log }_2}\left( {1 + {\gamma _k}} \right)}, \\
s.t.\quad \sum\limits_{k = 1}^K {{{\left\| {{{\bf{w}}_{k}}} \right\|}^2}}  &\le {P_{l,\max }},\quad \forall k \in K,l \in L,\\
\quad \quad {\theta _{r,n}} &\in \left[ {0,2\pi } \right],\quad \forall r \in R,n \in N,
\end{aligned}
\end{equation}
where the objective function pertains to an optimal problem with a multi-timescale horizon, $P_{l,max}$ denotes the maximum transmit power of the AP \textit{l}, and $\theta_{rn}$ denotes the reflection coefficient at the RISs, respectively. 

Given the complex characteristics of the non-convex objective function (9), the concurrent optimization of both the phase shift matrix and the precoding matrix presents a significant challenge. Nevertheless, drawing inspiration from MARL, we have proposed an innovative joint precoding network to tackle the optimization problem $\mathcal{P}^{\circ}$, as detailed in Section III.

\section{Proposed Joint Precoding and Phase Shift Optimization Framework}\label{se:framework}

\subsection{Overview of the Framework}
Within the context of a multi-agent architecture, every individual agent is constituted by two discrete components: an actor, which is responsible for the execution of actions, and a critic, which plays a pivotal role in the evaluation and refinement of the policy, respectively. MARL with centralized training decentralized execution (CTDE) has emerged as a viable alternative, streamlining centralized learning to a more computationally manageable extent. Consequently, the optimization problem denoted as $\mathcal{P}^{0}$ in equation (9) can be reformulated within the framework of MARL-CTDE as follows:

\begin{equation}
    \begin{aligned}
    \mathcal{P}^{1} \mathop {\max }\limits_{{{\bf{w}}_{lk}},\Theta } \text{sum-SE} \!&=\!\! \mathop \sum \limits_{k = 1}^K {\log _2}\!\left(\!1 \!+\! \frac{{{{\left| {\sum\limits_{l = 1}^L {{{\hat{\bf{h}}}_{lk,d}^{\rm{H}}} {{\bf{w}}_{lk}}} } \right|}^2}}}{{\sum\limits_{i = 1,i \ne k}^K {{{\left| {\sum\limits_{l = 1}^L { {{\hat{\bf{h}}}_{lk}^{\rm{H}}} {{\bf{w}}_{li}}}}\right|}^2}}\!\!+\!{\sigma ^2}}}\right),\\
    s.t.\quad \sum\limits_{k = 1}^K {{{\left\| {{{\bf{w}}_{lk}}} \right\|}^2}}  &\le {P_{l,\max }},\quad \forall k \in K,l \in L,\\
    \quad \quad {\theta _{r,n}} &\in \left[ {0,2\pi } \right],\quad \forall r \in R,n \in N.
\end{aligned}
\end{equation}
We transfer $\mathcal{P}^0$ to $\mathcal{P}^1$ for convenience in a MARL scenario. Each AP \textit{l}, instead of UE \textit{k}, is considered an agent, which is discussed in the following subsection.

\subsection{Fuzzy Logic}

Given the MARL policy in our network, it is evident that the RIS-aided CF mMIMO system, with a substantial number of APs and users, results in a large matrix calculation dimension and heightened complexity. Consequently, the conventional MARL necessitates simplification to ensure real-time interaction capability and scalability of the algorithms developed. Motivated by the integration of FL in a seminal work \cite{9764374}, we propose an innovative two-layer MARL-based downlink joint precoding method. This method strategically employs FL to formulate a correlation between fuzzy agents and entities, whose network is shown in Fig.\ref{Fig1}.

The precoding problem $\mathcal{P}^{1}$ and FL are described in this case, along with a MARL tuple $<\mathcal{S}^{(t)},\mathcal{A}^{(t)},r^{(t)}>$ at slot \textit{t}. The state space $\mathcal{S}^{(t)}= \left( \mathcal{S}_{1}^{(t)},\dots,\mathcal{S}_{n}^{(t)} \right)$, action space $\mathcal{A}^{(t)}= \left( \mathcal{A}_{1}^{(t)},\dots,\mathcal{A}_{n}^{(t)} \right)$, and reward $r^{(t)}=\left( r_{1}^{(t)},\dots,r_{n}^{(t)} \right)$ are designed as follows.

\subsubsection{Agent} 
We consider each AP as an agent.
\subsubsection{State space}
States are characterized as comprehensive representations of the entire system. To encompass the states of UEs dispersed across diverse locations, we employ an observational approach comprising both partial state variables and global state variables. In this framework, the agent is equipped to observe the relative positions of all UEs concerning all APs, denoted as \textbf{\textit{D}}. Specifically, local state is contemplated for AP \textit{l} at slot \textit{t}.
\begin{align}
    \mathcal{S}_{l}^{(t)} = \left(\mathbf{H}_{l,k}^{(t)}, \mathbf{F}_{r,k}^{(t)}, \mathbf{\Theta}_{r}^{(t)},\mathbf{G}_{l,r}^{(t)},\textbf{\textit{D}}_{l},\mathbf{w}_{l}^{(t)},\Theta^{(t)}, \gamma_{l,k}^{(t)} \right). 
\end{align}

\subsubsection{Initialization}
Initially, every fuzzy agent's fuzzy state is denoted as $\hat{\mathcal{S}}^{(t)} = \left( \hat{\mathcal{S}}_{1}^{(t)},\dots,\hat{\mathcal{S}}_{n}^{(t)} \right)$, with each $\mathcal{S}_{i}^{(t)}$ being a random selection from the observed state, and $n$ representing the total number of fuzzy agents. Subsequently, we divide each dimension of the state space into $n$ unique fuzzy sets. For any given $j$-th dimension, the fuzzy state set is expressed as $\left(\hat{s}_{1,j}^{(t)},\hat{s}_{2,j}^{(t)},\dots,\hat{s}_{K,j}^{(t)}\right)$. The corresponding membership function is denoted as $\xi_{\hat{s}_{i,j}(s)}^{(t)} = \mathbf{exp}(-\frac{1}{d_{a}*n}|s^{(t)}-\hat{s}_{i,j}^{(t)}|)$,where $d_{a}$ represents the dimension of the action space \cite{Liu2023DoubleLayerPC}.

\subsubsection{Fuzzy Action space}
Each fuzzy agent in a fuzzy system is assigned a policy based on the perceived fuzzy state, which is represented as $\mathcal{S}^{(t)}$. Subsequently, defuzzification is employed to establish a mapping from the fuzzy action $\hat{\mathcal{A}}^{(t)} = \left( \hat{\mathcal{A}}_{1}^{(t)}, \dots, \hat{\mathcal{A}}_{n}^{(t)} \right)$ to the specific action $\mathcal{A}^{(t)}$. The mapping relationship between the $p$-th agent and the $i$-th fuzzy agent is denoted as $\Xi_{i,p}^{(t)} = \prod_{i=j}^{d_{a}} \xi_{\hat{s}_{i,j}\left(s_{k,j}\right)}^{(t)}$. Here we define the relationship as $\mathcal{A}_{p}^{(t)}=\sum_{i=1}^{m} \bar{\Xi}_{i,p}^{(t)} \times \hat{\mathcal{A}}_{i}^{(t)}$, where $\bar{\Xi}_{i,p}$ represents the normalized mapping relationship.

\subsubsection{Reward space}
Subsequent to the reception of the specific action $\mathcal{A}^{(t)}$ by the agents, the corresponding reward $r$ is ascertained in accordance with the predefined reward function. In the context of interacting with the environment, it is essential to employ fuzzy agents rather than entities. This necessitates the process of fuzzification to derive the fuzzy reward $\hat{r}^{(t)}=\left( \hat{r}_{1}^{(t)},\dots,\hat{r}_{n}^{(t)} \right)$ within the framework of reinforcement learning. The fuzzy reward can be expressed as $\hat{r}_{i}^{(t)}=\sum_{k=1}^{K} \bar{\Xi}_{i,k}^{(t)} \times r_{k}^{(t)}$. Therefore, the incorporation of fuzzification is imperative for the successful completion of the reinforcement learning model.

\subsection{FL-MARL Algorithm}
In the context of FL-MARL, individual fuzzy agents independently calculate the policy gradient for their respective local actor networks, utilizing the collective abstract state and action as a basis. Additionally, the objective function for the $i$-th policy, denoted as $\pi_{i}$, can be formulated as $L(\pi_{i}) = \sum_{\hat{\mathcal{S}}_{i}^{(t)}} p_{\pi}\left(\hat{\mathcal{S}}_{i}^{(t)}\right) \sum_{\hat{\mathcal{A}}_{i}^{(t)}} \pi\left(\hat{\mathcal{A}}_{i}^{(t)} \mid \hat{\mathcal{S}}_{i}^{(t)}\right) \hat{r}_{i}^{(t)}$.

\begin{algorithm}[!t]
    \caption{FL-MARL Algorithm for Maximizing sum-SE}
    \label{alg:AOA}
    \renewcommand{\algorithmicrequire}{\textbf{Input:}}
    \renewcommand{\algorithmicensure}{\textbf{Output:}}
    \begin{algorithmic}[1]
    \State \textbf{Initialize} AP agent states $\mathcal{S}_{1}^{(t)},\dots,\mathcal{S}_{n}^{(t)}$ by randomly sampling fuzzy agent states: $\hat{\mathcal{S}}_{1}^{(t)},\dots,\hat{\mathcal{S}}_{n}^{(t)}$ 
    \State count = 0
    \While {count $\le$ N} 
        \State Evaluate the network actor to decide the downlink precoding and phase shift design: $\hat{\mathcal{A}}_{i}^{(t)} = \pi_{i}\left(\mathcal{S}_{i}^{(t)}\right)$ 
        \State Calculate actual actions $\mathcal{A}_{i}^{(t)} = \sum_{i=1}^{m}\bar\Xi_{i,k}^{(t)} \times \hat{\mathcal{A}}_{i}^{(t)} $
        \State Get actual rewards $r_{i}$ with reward function
        \State Use fuzz function: $\hat{r}_{i}^{(t)}=\sum_{k=1}^{K} \bar{\Xi}_{i,k}^{(t)} \times r_{k}^{(t)}$ to calculate fuzzy rewards $\hat{r}_{i}$
        \State Update environment
        \State Get next actual states $\mathcal{S}_{i}^{(t)}$ 
        \State Get next fuzzy states $\hat{\mathcal{S}}_{1}^{(t)}$ by fuzz function: $\hat{\mathcal{S}}_{i}^{(t)} = \sum_{k=1}^{K}\bar\Xi_{i,k}^{(t)} \times \hat{\mathcal{S}}_{k}^{(t)}$
        \State Update membership function with $\xi_{\hat{s}_{i,j}\left(\hat{s}_{k,j}\right)}^{(t+1)}$
        \State Store fuzzy experience $<\mathcal{S}^{(t)},\mathcal{A}^{(t)},r^{(t)}>$ in replay buffer $\mathcal{D}_{i}$
        
        \For {each training step}
            \State Randomly sample a mini-batch of $\mathcal{B}_{i}$ transitions uniformly from $\mathcal{D}_{i}$
            \State Update weights of joint precoding and phase shift critic network
            \State Calculate policy gradient of the two layer actor network $\Delta J(\theta_{Q_{\pi}})$ and update target network
        \EndFor
        \State count += 1
    \EndWhile
    \end{algorithmic}

\end{algorithm}

The $i$-th fuzzy reward, symbolized as $r_{i}$, is correspondingly linked with a universal fuzzy action, denoted as $\mathcal{A}$, and a state, represented as $\mathcal{S}$. This association results in a consolidated global behavior value, $Q_{\pi} \left (\hat{\mathcal{S}}^{(t)}, \hat{\mathcal{A}}^{(t)}\right)$, which is the output of the $i$-th critic network's computation. In the context of $\pi_{i}$, the policy gradient of the local actor network can be articulated as follows:
\begin{align}
\Delta \!J\!\left(\theta_{\pi_{i}}\right) \!\!=\!\!\! \sum^{\hat{\mathcal{A}}_{i}^{(t)}}\!\! Q_{\pi}\!\left(\hat{\mathcal{S}}^{(t)}, \hat{\mathcal{A}}^{(t)}\!\!\right)\! \Delta\!\pi_{i}\!\left(\hat{\mathcal{A}}_{i}^{(t)} \!\!\mid \!\!\hat{\mathcal{S}}_{i}^{(t)} ; \!\theta_{\pi_{i}}\!\right),  
\end{align}
where $\Delta\pi_{i}\left(\hat{\mathcal{A}}_{i}^{(t)} \mid \hat{\mathcal{S}}_{i}^{(t)} ; \theta_{\pi_{i}}\right) $ is the output by the local policy network.

In accordance with Algorithm 1, the soft update procedure is implemented concurrently with the existing network configuration. The target actor network undergoes modification according to the expression $\theta_{\pi_{i}^{\prime}} \leftarrow \tau \theta_{\pi_{i}^{\prime}} + (1-\tau) \theta_{\pi_{i}}$, whereas the target critic network is adjusted based on the formula $\theta_{Q_{\pi^{\prime}}} \leftarrow \tau \theta_{Q_{\pi^{\prime}}} + (1-\tau) \theta_{Q_{\pi}}$.

\subsection{Complexity comparison}

We compare the computational complexity of different algorithms. We assume that $Q_a$ and $Q_c$ denote the output size of the $a$-th and $c$-th layer or the input size of the next layer, and $Q_{SE}$ represents the computational complexity of calculating SE expressions, respectively. Note that as observed from Table I, the computational complexity of MADDPG exhibits an exponential increase with the number of APs $L$ increasing. In contrast, upon integrating FL, the computational complexity increases linearly with the number of APs $L$. Correspondingly, the introduction of FL makes the computational complexity of FL-MADDPG linearly related to the number of fuzzy agents $N_{F}$, thus reducing the need for high backhaul capacity for ${\left(\frac{N_{F}}{L}\right)}^{2}$.

\begin{table*}[t]
\caption{Comparison of Computational Complexity.}
	\label{tab1}
	\centering
	\footnotesize
	\renewcommand{\arraystretch}{1.5}
\begin{tabular}{ll}
    \hline\hline
    \textbf{\makecell{Parameters}} &\textbf{\makecell{Computational Complexity}} \\ \hline\hline 
    \emph{\makecell{\textbf{MADDPG}}} & \makecell{$\left\{\mathcal{O}(L^2MKN^2\sum_{a=1}^{A_L}Q_a^{2}\!\!+\!\!L^2MK\sum_{c=1}^{C_L}Q_c^{2})\right\} \times \left\{\mathcal{O}(L^2MKN^2\sum_{a=1}^{A_H}Q_a^{2}\!\!+\!\!LN\sum_{c=1}^{C_H}Q_c^{2} \!\!+\!\! L^3Q_{SE})\right\}$} \\ \hline
    \emph{\makecell{\textbf{FL-MADDPG}}} &  \emph{\makecell{$\left\{\mathcal{O}(LN_FMKN^2\sum_{a=1}^{A_L}Q_a^{2}\!\!+\!\!LN_FMK\sum_{c=1}^{C_L}Q_c^{2})\right\}\times \left\{\mathcal{O}(LN_FMKN^2\sum_{a=1}^{A_H}Q_a^{2}\!\!+\!\!LN\sum_{c=1}^{C_H}Q_c^{2} \!\!+\!\! L^2N_FQ_{SE})\right\}$}} \\ \hline\hline
\end{tabular}
\end{table*}

\section{Simulation Results}\label{se:simulation}


\begin{figure}[t]
    \centering
    \includegraphics[width=0.4\textwidth]{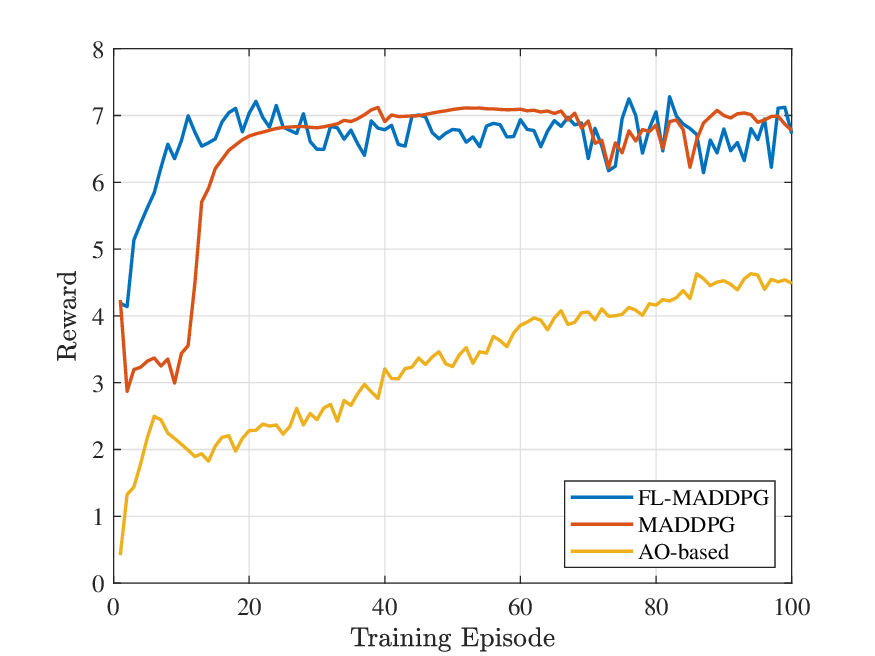}
    \caption{Average reward against the training step with $step$ = 100, $L$ = 4, $K$ = 4, $R$ = 4, $M$ = 8, $U$ = 1, and $N$ = 16.}
    \label{Fig2} 
\end{figure}

\begin{figure}[t]
    \centering
    \includegraphics[width=0.4\textwidth]{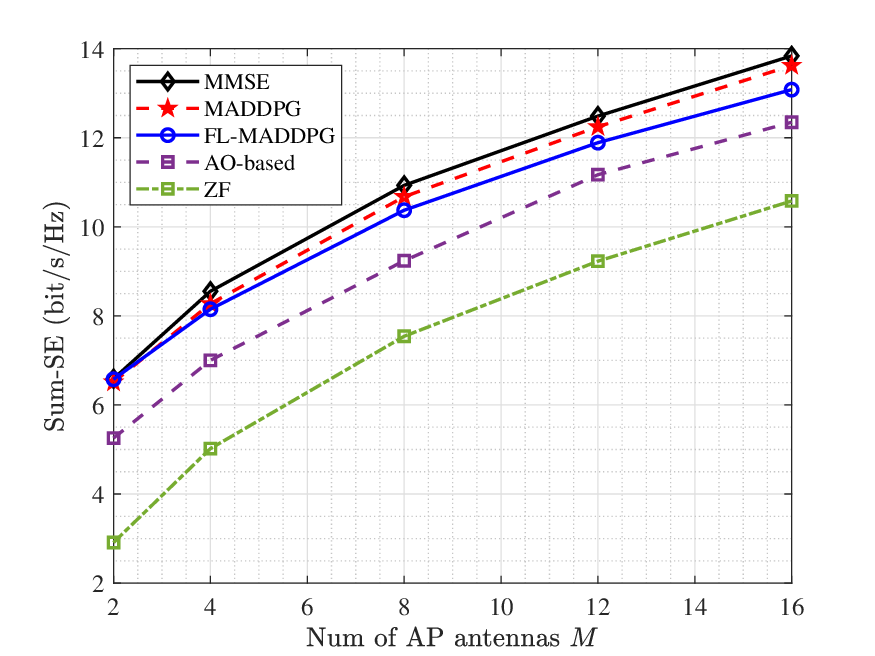}
    \caption{Sum-SE against the number of AP antennas with $L$ = 4, $K$ = 4, $R$ = 4, $U$ = 1, and $N$ = 16.}
    \label{Fig3} 
\end{figure}

\begin{figure}[t]
    \centering
    \includegraphics[width=0.4\textwidth]{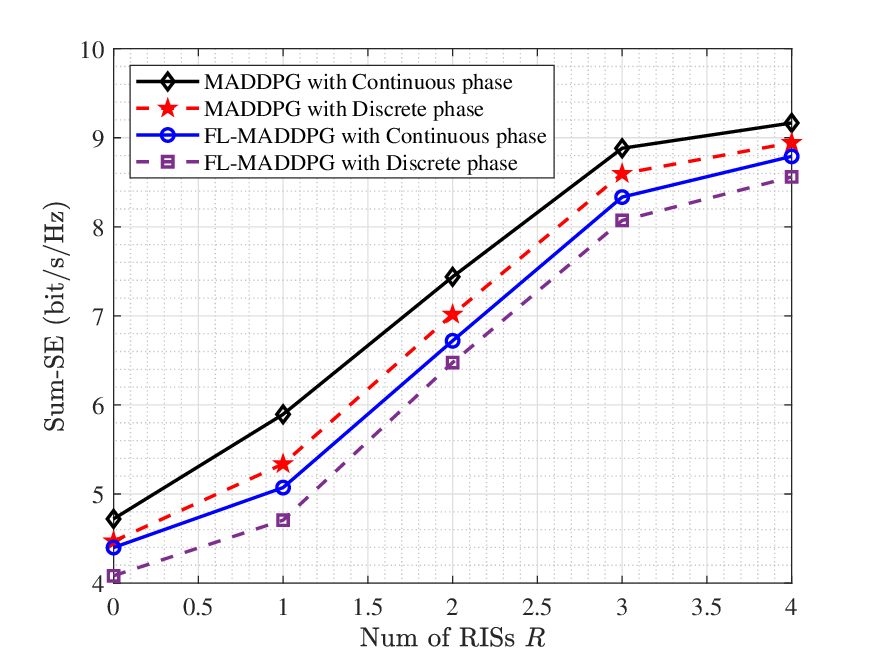}
    \caption{Sum-SE against the number of RISs with $L$ = 4, $K$ = 4, $M$ = 8, $U$ = 1 and $N$ = 16.}
    \label{Fig4} 
\end{figure}

\begin{figure}[t]
    \centering
    \includegraphics[width=0.4\textwidth]{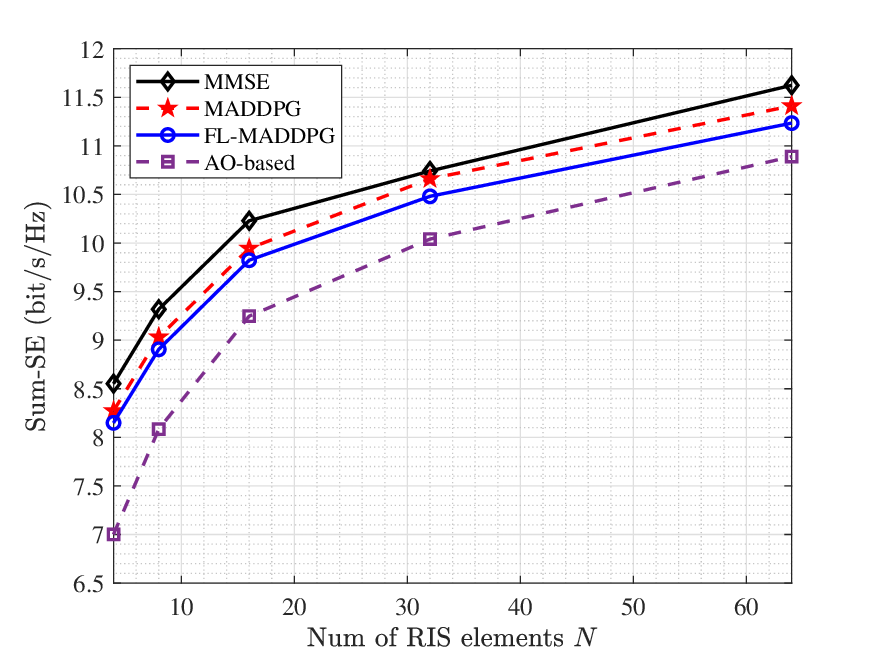}
    \caption{Sum-SE against the number of RIS elements with $L$ = 4, $K$ = 4, $R$ = 4, $M$ = 8, and $U$ = 1.}
    \label{Fig5} 
\end{figure}

\subsection{Simulation Setup}

In the proposed RIS-aided cell-free network simulation, we consider a 50 m $\times$ 50 m region served by a cell-free network with all APs simultaneously serving all UEs. Each AP divides the area into four equal squares, and UEs are randomly deployed within these squares. To enhance network capacity, RISs are strategically placed at the center of each of the four equal squares that the APs divide the area into. We assume that the maximum transmit power for APs is $P_{l,\text{max}} = 0$ dBm, and the initial number of antennas per AP is $M = 4$, with a noise power of $\delta^{2} = -96$ dBm. Considering the limited antennas and low transmit power of APs in cell-free networks, we adopt the channel model from \cite{9110869}. For the performance enhancement of sum SE, we consider a proper experience pool size in the simulation to improve the generalization ability of the model. The experimental details of MARL are given in Table II.

\begin{table}[t]
\centering
    \fontsize{9}{12.5}\selectfont
    \caption{The Model Structure and Experimental Details.}
    \label{Table1}
    \begin{tabular}{ccc}
    \toprule
    \bf Parameters &  \bf Size \\
    \midrule
    Hidden layer of AP Precoding & 512, Leaky Relu (0.01)\\
    Hidden layer of RIS Phase Shift & 256, Leaky Relu (0.01) \\
    Mini-batch & 32 \\
    Discounted factor $\gamma^{\mathrm{p}}$ and $\gamma^{\mathrm{f}}$ & 0.99 \\
    Experience pool size $\mathcal{D}^{\mathrm{p}}$ and $\mathcal{D}^{\mathrm{f}}$ & 4096 and 2048 \\
    Soft update rate $\tau^{\mathrm{p}}$ and $\tau^{\mathrm{f}}$ & 0.0001 and 0.001 \\
    \bottomrule
    \end{tabular}\vspace{-0.2cm}
\end{table}



\subsection{Convergence of the proposed algorithm}
To demonstrate the convergence of the proposed algorithms, we present a plot of the reward as a function of training time in Fig. 2. The outcomes depicted in Fig. 2 show that the MARL method can converge faster in the same number of training steps than the traditional AO-based method \cite{gan2022tcomm}. It is worth noting that in the initial stage of algorithm training, MADDPG showed a more stable upward trend compared with FL-MADDPG. However, FL-MADDPG has a faster convergence rate, about 30\% faster than MADDPG, allowing FL-MADDPG to reach a stable state earlier due to its combination of fuzzy logic mechanisms to complete the agent-to-fuzzy agent mapping. Therefore, FL-MADDPG is more stable in comparison, demonstrating that the integration of FL into MADDPG yields substantial savings in computing resources.

\subsection{Impact of key system parameters} 
We evaluate the sum-SE of the proposed RIS-aided cell-free network in this subsection.

\subsubsection{SE against the number of antennas per AP} 
We illustrate the average sum-SE in relation to the number of AP antennas in Fig. 3. The figure reveals a notable increase in the sum-SE across all instances as the number of AP antennas increases. Notably, FL-MADDPG demonstrates a performance closely aligned with MADDPG, which is near MMSE. SE has a 42\% gain compared to ZF and 18\% over the AO-based algorithm, signifying that our proposed FL-MADDPG framework can approximate the performance of MADDPG with a relatively fast convergence.

\subsubsection{SE against the number of RISs} 
We depict the sum-SE relative to the number of RISs in Fig. 4. It is easy to find that with the increase in RIS, the sum-SE shows an increasing trend. The gap between the continuous phase and the discrete phase gradually decreases with the increase in the number of RISs, which is because there are blind areas in CF-mMIMO. These blind areas need RIS to provide and enhance communication services. It is worth noting that the percentage gap between FL-MADDPG and MADDPG under continuous is smaller than the gap under discrete, indicating that FL-MADDPG is more suitable for a broader continuous scenario.

\subsubsection{SE against the number of RIS elements} 
We depict the sum-SE relative to the number of RIS elements in Fig. 5. As the number of RIS elements increases, the sum SE of the RIS-aided cell-free network shows a notable improvement. Compared with alternate optimization, the MARL we adopted has more than 10\% performance improvement, which can quickly approach the theoretical value in a limited training time. However, this enhancement is accompanied by heightened implementation complexity of the RIS. Furthermore, this state underscores the robustness of our proposed algorithm across a broader spectrum of application scenarios, approaching performance levels indicative of optimality.

\section{Conclusions}\label{se:conclusion}In this paper, we investigated the maximization of downlink SE in a RIS-aided CF mMIMO system through joint precoding and phase shift design. To achieve this goal, we proposed a MARL-based method incorporating fuzzy logic. The method presented leverages parallel computing to diminish computational time, rendering it highly suitable for deployment in expansive networks. Our simulation findings substantiate that the MARL method, underpinned by fuzzy logic, significantly curtails computational complexity by approximately 42\%, thereby enhancing reliability in practical settings relative to traditional AO-based algorithms in a limited training time. In future work, it is interesting to investigate the simultaneous transmitting and reflecting (STAR)-RIS-aided CF mMIMO systems and design joint precoding with imperfect CSI to enhance the network.

\bibliographystyle{IEEEtran} \bibliography{IEEEabrv,Bibliography}

\end{document}